\documentclass[12pt]{article}
\usepackage{graphicx}
\usepackage{amsfonts}
\usepackage{amssymb}
\usepackage{latexsym}
\usepackage{color}
\usepackage{cite}
\input{colordvi.tex}

\renewcommand{\thefootnote}{\#\arabic{footnote}}

\newcommand{\bea}{\begin{eqnarray}}  \newcommand{\eea}{\end{eqnarray}}
\newcommand{\beq}{\begin{equation}}  \newcommand{\eeq}{\end{equation}}

\begin{document}

\setcounter{footnote}{0}
\begin{titlepage}

\begin{center}

\hfill May 2009\\

\vskip .5in

{\Large \bf Gravitino Dark Matter and Non-Gaussianity}

\vskip .45in

{\large
Tomo Takahashi$^1$, 
Masahide Yamaguchi$^2$,\\ 
\vspace{0.3cm}
Jun'ichi Yokoyama$^{3,4}$ and Shuichiro Yokoyama$^5$
}

\vskip .45in

{\em
$^1$
Department of Physics, Saga University, Saga 840-8502, Japan \\
$^2$Department of Physics and Mathematics, Aoyama Gakuin
University, Sagamihara 229-8558, Japan \\
$^3$Research Center for the Early Universe (RESCEU), Graduate
School of Science, The University of Tokyo, Tokyo 113-0033, Japan \\
$^4$Institute for the Physics and Mathematics of the Universe (IPMU),
The University of Tokyo, Kashiwa, Chiba, 277-8568, Japan\\
$^5$Department of Physics and Astrophysics, Nagoya University,
Aichi 464-8602, Japan
}

\end{center}

\vskip .4in

\begin{abstract}
We investigate density fluctuations in a scenario with gravitino dark
matter in the framework of modulated reheating, which is
known to generate large non-Gaussianity. 
We show that gravitino dark
matter is disfavored in this framework.
We also briefly discuss
  the case with  the curvaton mechanism and some other possible dark matter scenarios.
\end{abstract}

\end{titlepage}

\renewcommand{\thepage}{\arabic{page}}
\setcounter{page}{1}
\renewcommand{\thefootnote}{\#\arabic{footnote}}

Recent progress in high-precision observational cosmology with 
cosmic microwave background (CMB) radiation has motivated us to study the
tiny nonlinear effects in the very early stage of the Universe when the
amplitude of fluctuations was much smaller than unity.  Among various
observables to quantify non-Gaussianity of primordial fluctuations, the
so-called non-linearity parameter $f_{\rm NL}$, which is a stunted
parameter to quantify the degrees of freedom of actual
non-Gaussianity, has been widely discussed recently because observations
of CMB is becoming to be able to measure its possible deviation from
zero.  From the recent result from WMAP5, the so-called local-type
non-linearity parameter $f_{\rm NL}$ is constrained as 
$ -9 < f_{\rm NL}
< 111$ \cite{Komatsu:2008hk} and $ -4 < f_{\rm NL}< 80$
\cite{Smith:2009jr} at 95 \% confidence level (CL).
Although purely Gaussian fluctuation with $f_{\rm
NL}=0$ is perfectly consistent with the data, its observed central value
is so much deviated from zero that future higher-precision observations
may well confirm this local-type non-Gaussianity.  Then the simplest
models of single-field inflation \cite{inf} would be ruled out as a
generation mechanism of primordial density/curvature fluctuations
\cite{yuragi}.

We can think of two mechanisms as an alternative to the inflaton to
generate primordial fluctuations, modulated reheating
\cite{Dvali:2003em} and the curvaton
\cite{Mollerach:1989hu,Enqvist:2001zp}, which are motivated in particle
physics because there are many light fields in generic supersymmetric
theories.  Neither of them is necessary as long as the inflaton can be
responsible for the generation of primordial fluctuations.  But if the
presence of large non-Gaussianity was established, they would be a
target of serious study\footnote{
Other phenomenological generation
mechanisms of large local-type non-Gaussianity include Refs.\
\cite{others}.
}.  In fact, non-Gaussianity in these alternative
scenarios has been studied already in \cite{Zaldarriaga:2003my,Suyama:2007bg} for
modulated reheating and in \cite{Lyth:2002my,Bartolo:2003jx} for the
curvaton, and it has been shown that in both scenarios large local-type
non-Gaussianity is possible depending on the values of model parameters
and/or initial conditions.

The purpose of this Letter is to argue that confirmation of
non-Gaussianity of local-type would not only rule out the simplest
single-field slow-roll inflation models but also constrain the identity
of the dark matter. That is, we show that the gravitino dark matter
would confront difficulty in the two representative generation
mechanisms of fluctuations mentioned above.  Before proceeding to this
main topic, for completeness, let us mention the relevance of the other
type of non-Gaussianity, namely the equilateral-type, to the nature of
the dark matter.  It is generated due to nonlinear interactions during
inflation.  Typical models generating such non-Gaussianity are those
making use of scalar fields with non-canonical kinetic terms~\cite{equil}. They do
not give us any information on the nature of the dark matter, because
the inflaton is responsible for the fluctuation in this case.
Observationally, the non-linearity parameter of the equilateral-type,
$f_{\rm NL}^{\rm equil}$, has been much less constrained than the local-type,
and as long as the former is concerned, there is no hint of
non-Gaussianity so far.  We therefore concentrate on the local-type
non-Gaussianity which we denote simply by $f_{\rm NL}$ as above.

In models with a supersymmetric extension of the standard model, the
gravitino, the spin-$3/2$ superpartner of the graviton, arises when the
global supersymmetry is promoted to a local symmetry.  If it is the
lightest supersymmetric particle (LSP) and stable because of the
$R$-parity conservation, it can be a good candidate of dark matter (DM)
of the Universe.  Gravitinos can be produced in scattering processes of
particles in the thermal bath and its thermal relic density depends on
the efficiency of the production, which is sensitive to the reheating
temperature $T_R$ after inflation. Its relic abundance is given as
$Y_{3/2} \equiv n_{3/2} / s \propto T_R$ where $Y_{3/2}$ is the yield
from the thermal production, $n_{3/2}$ and $s$ are the number density of
gravitinos and the entropy of the universe \cite{Ellis:1984eq}.
Recently, it was pointed out that gravitinos are also produced
non-thermally from the decay of heavy scalar fields like an inflaton or
moduli if they acquire non-vanishing expectation values
\cite{Nakamura:2006uc}. The abundance of gravitinos produced
non-thermally also depends on the reheating temperature, in fact, is
proportional to the inverse of the reheating temperature. Then, when one
considers the modulated reheating scenario as a generation mechanism of
primordial fluctuations, the dependence on $T_R$ would result in the
generation of isocurvature fluctuations in gravitino DM since the
modulated reheating scenario indicates that $\delta T_R / T_R \ne
0$. Such isocurvature fluctuations can be strongly constrained from
cosmological observations such as CMB. From WMAP5, the ratio of
isocurvature fluctuations to curvature fluctuations, which is denoted by
$\alpha$, is constrained as $\alpha < 0.072$ (axion type; uncorrelated)
and $\alpha < 0.0041$ (curvaton type; anti-correlated) for WMAP+BAO+SN
at the $95\%$ CL \cite{Komatsu:2008hk} (see also
Ref.~\cite{Hikage:2008sk} for the constraint on isocurvature
fluctuations from the non-Gaussianity).

One may classify the origin of gravitino DM to the following two cases
in the context of the modulated or the curvaton scenario. The first one
is that gravitino can be produced at the decay of the inflaton through
the rescattering process of thermal plasma (thermal production) or
directly by the inflaton decay (non-thermal production). In this case,
such gravitino DM results in the generation of large isocurvature
fluctuations not only in the modulated scenario but also in the curvaton
scenario \cite{Lyth:2002my}. The other one is the case that the curvaton
mechanism is operative and gravitino DM is mostly produced from the
decay of the curvaton. In this case, isocurvature fluctuations are suppressed if and
only if the curvaton practically dominates the energy density of the
Universe just before its decay \cite{Lyth:2002my}. However, such a case
cannot generate the large non-Gaussianity. Thus when large
non-Gaussianity is confirmed in the future, a scenario with gravitino DM
may not be a viable candidate as the main constituent of the DM in the
Universe.

In this Letter, we discuss in detail the viability of the gravitino DM
scenario when non-Gaussianity is large, particularly focusing on the
modulated reheating scenario \cite{Dvali:2003em} since the issues of
isocurvature fluctuations in this scenario is relatively unexplored.  As
shown below, the gravitino DM acquires large isocurvature fluctuations in
this scenario, which is too large to be consistent with current
observations of CMB.

Now we briefly review the non-linearity parameter which characterizes
non-Gaussianity of the fluctuations.  For this purpose, it is
fashionable to write the comoving curvature perturbation, $\zeta$, up to
the second order to discuss the bispectrum as
\begin{equation}
\label{eq:zeta_fNL}
\zeta = 
\zeta_{(1)}
+ \frac{3}{5} f_{\rm NL} \zeta_{(1)}^2~,
\end{equation}
where $\zeta_{(1)}$ is the Gaussian part of curvature perturbation and
$f_{\rm NL}$ characterizes the size of non-Gaussianity observable in the
bispectrum.  With this parameterization, the definition of the bispectrum $B_\zeta$ 
is given by 
\begin{eqnarray}
  \langle \zeta_{\vec k_1} \zeta_{\vec k_2} \zeta_{\vec k_3} \rangle
  &=&
  {(2\pi)}^3 B_\zeta (k_1,k_2,k_3) 
  \delta ({\vec k_1}+{\vec k_2}+{\vec k_3}). \label{eq:bi}
\end{eqnarray}
The leading contribution of the bispectrum $B_\zeta$  can be written as
\begin{eqnarray}
B_\zeta (k_1,k_2,k_3)
&=&
\frac{6}{5} f_{\rm NL} 
\biggl( 
P_\zeta (k_1) P_\zeta (k_2) 
+ P_\zeta (k_2) P_\zeta (k_3) 
+ P_\zeta (k_3) P_\zeta (k_1)
\biggr). \label{eq:bi_zeta}
\end{eqnarray}
Here $P_\zeta$ is the power spectrum which is defined as
\begin{eqnarray}
  \label{eq:power}
  \langle \zeta_{\vec k_1} \zeta_{\vec k_2} \rangle
  =
  {(2\pi)}^3 P_\zeta (k_1) \delta ({\vec k_1}+{\vec k_2}).
\end{eqnarray}

In order to evaluate the curvature perturbations $\zeta$, we adopt the
$\delta N$ formalism~\cite{Starobinsky:1986fxa}.  The curvature
perturbation on sufficiently large scales $\zeta$ at time $t_f$ is equal
to the perturbation in time integral of the local expansion from an
initial flat hypersurface ($t=t_i$) to the final uniform energy density
hypersurface. Since the local expansion on sufficiently large scales can
be approximated by the expansion of the unperturbed Friedmann universe,
the curvature perturbation $\zeta$ is evaluated as
\begin{eqnarray}
\zeta (t_f, {\vec x}) = N(t_i,t_f,{\vec x})-({\rm spatial~average}), \label{delN1}
\end{eqnarray}
where the number of $e$-folds $N(t_i,t_f,{\vec x})$ is defined by the
time integral of the local Hubble parameter as,
\begin{eqnarray}
N(t_i,t_f,{\vec x})=\int_{t_i}^{t_f} H(t,{\vec x})dt. \label{delN2}
\end{eqnarray}
In the $\delta N$ formalism, taking $t_i=t_*$ which is a certain time
well after the relevant scale crossed the horizon scale, the curvature
fluctuations from fluctuations of scalar fields are given as,
\begin{eqnarray}
  \zeta = N_a \delta \varphi^a_\ast
  +\frac{1}{2}N_{ab} \delta \varphi_*^a \delta \varphi_*^b
  + \cdots,
  \label{eq:zeta_expansion}
\end{eqnarray}
up to the second order, where $\delta\varphi^a_*$ is the long-wave
frozen part of the fluctuation of the scalar field $\varphi^a$ on the
initial flat hypersurface at $t=t_*$ and $N_a$ and $N_{ab}$ are given by
\begin{eqnarray}
  N_a \equiv \frac{\partial N}{\partial \varphi^a}, ~~~
  N_{ab} \equiv 
  \frac{\partial^2 N}{\partial \varphi^a \partial \varphi^b}.
\end{eqnarray}
Here the summation is implied for the repeated indices.

The non-linearity parameter is then given by
\begin{eqnarray}
{6 \over 5}f_{NL} &\!=\!& {N_{ab}N^aN^b \over \left(N_cN^c\right)^2}~,\label{fnlbase}\\
\end{eqnarray}
where the indices are lowered and raised by using the Kronecker's delta
$\delta^{ab}$.

Using $\delta N$ formalism, we can evaluate the curvature perturbations
in the modulated reheating scenario.  In this scenario, fluctuations,
$\delta \sigma$, of some modulus, $\sigma$, renders fluctuations in the
decay rate of the inflaton.  To evaluate $\zeta$ from the fluctuations
of the modulus, we need to follow the background evolution from the
epoch after the inflaton begins to oscillate around the minimum of the
potential whose shape is assumed to be quadratic below.  Then the energy
density of the oscillating inflaton field behaves like matter, thus the
background equations after inflation are governed by the following
equations:
\begin{eqnarray}
&&\frac{d \rho_{\phi}}{dN}+3 \rho_{\phi}=-\frac{\Gamma}{H} \rho_{\phi}, \label{bd1} \\
&&\frac{d \rho_r}{dN}+4 \rho_r=\frac{\Gamma}{H} \rho_{\phi}, \label{bd2} \\
&&H^2=\frac{1}{3} (\rho_{\phi}+\rho_r), \label{bd3}
\end{eqnarray}
where $\Gamma$ is the $\sigma$-dependent 
decay rate of the inflaton and $\rho_{\phi}$ and
$\rho_r$ are energy densities of inflaton $\phi$ and radiation,
respectively. We assume the energy density of the modulus is negligible
throughout. The reduced Planck scale is set to be unity here.

By solving the above equations from the end of inflation to the
completion of reheating with the initial condition
$\rho_{\phi}(0)=\rho_0=3H_0^2$ and $\rho_r(0)=0$, 
we can obtain a relation
between $N$ and $\Gamma$.  The number of $e$-folds, $N$, elapsed
while the
Hubble parameter drops from $H_0$ 
to $H_f$ ($H_f \ll \Gamma$) can be written
formally as \cite{Suyama:2007bg}
\begin{eqnarray}
N =\frac{1}{2} \log \frac{H_0}{H_f}+Q \left( \frac{\Gamma}{H_0} \right), \label{bd4}
\end{eqnarray}
where the function $Q$ is defined by 
\begin{eqnarray}
\exp \bigg[ 4Q ( \Gamma /H_0 ) \bigg] \equiv \int_0^\infty dN'~\frac{\Gamma}{H(N')}e^{4N'}\frac{\rho_\phi(N')}{\rho_0}. \label{bd5}
\end{eqnarray}
On dimensional grounds $Q$ depends only on the combination of $x \equiv
\Gamma/H_0$.  
Let us consider the typical case
 that the final stage of reheating is governed by a perturbative
decay of the inflaton after a certain period of field-oscillation 
dominated stage.  Then $x$ is much smaller than unity and we find
the following analytic form of $Q(x)$ \cite{Suyama:2007bg},\footnote{
  The behavior of the inflaton after inflation depends on the potential
  shape and the decay rate. Here we consider the case that the inflaton
  behaves like matter assuming the effective potential around the
  minimum is quadratic and oscillates enough before its decay. For other
  cases, some coefficients are changed but the essential results remain
  intact.
} 
\begin{eqnarray}
Q(x)=-\frac{1}{6} \log x+{\cal O}(x), ~~~~~x\ll 1. \label{bd7}
\end{eqnarray}

In the modulated reheating scenario, the curvature perturbation $\zeta$
is evaluated as\cite{Suyama:2007bg}
\begin{eqnarray}
\zeta (N_F) = Q_\sigma \delta \sigma_* + {1 \over 2} Q_{\sigma\sigma} \delta \sigma^2_*,
\label{modcurv}
\end{eqnarray}
up to the second order with
\begin{eqnarray}
Q_\sigma &\equiv& Q'(x) {\Gamma'(\sigma) \over H_0}~,\label{modfirst}\\
Q_{\sigma\sigma} &\equiv& Q'(x) {\Gamma''(\sigma) \over H_0} + Q''(x) \left({\Gamma'(\sigma) \over H_0}\right)^2~,
\label{modsecond}
\end{eqnarray}
where $Q'(x)=dQ(x)/dx$ and $\Gamma'(\sigma)=d\Gamma(\sigma)/d\sigma$.

From the above equations, we can obtain the following expression for the
non-linearity parameter
\begin{eqnarray}
{6 \over 5}f_{\rm NL} &\!\simeq\!& 
6 \left( 1 
- {\left(\Gamma''(\sigma)/\Gamma(\sigma)\right) \over
\left(\Gamma'(\sigma)/\Gamma(\sigma)\right)^2}\right) \label{modfnl}
\end{eqnarray}
in the case $x \ll 1$.

Hence, if we adopt proper form of the decay rate, we can realize a large
non-Gaussianity in the modulated reheating scenario. As an example, let
us consider the case that the decay rate depends on $\sigma$ as
\begin{eqnarray}
\Gamma (\sigma) = \Gamma_0 \left[ 1+A {\sigma \over M} + B \left({\sigma \over M}\right)^2 
\right]~,\label{simplemodel}
\end{eqnarray}
where $A$ and $B$ are some coefficients, $M$ is a some energy
scale and we have assumed $\sigma/M \ll 1$ and truncated at the second
order in $\sigma / M$.  In this model, we find an simple expression for $f_{\rm NL}$ as
\begin{eqnarray}
{6 \over 5}f_{\rm NL} &\! \simeq \!& 6
\left(1-{2B \over A^2}\right)~.\label{anafnl}
\end{eqnarray}
Thus, if we take $|B| / A^2 \gg 1$ and $B < 0$, we
can obtain a large and positive value of $f_{\rm NL}$.

Let us move on to the isocurvature fluctuations from gravitino DM in the
modulated reheating scenario.  As mentioned before, there are several
distinct ways to produce gravitinos.  One is the scattering of particles
in the thermal plasma and their relic abundance is evaluated as
\cite{Ellis:1984eq}
\begin{equation}
  Y_{3/2} \equiv
  \frac{n_{3/2}}{s} 
  \simeq
  10^{-12} \sum_i g_i^2
              \left(1 + \frac{m_{Gi}^2}{3 m_{3/2}^2}\right)
              \left(\frac{T_R}{10^{10}~{\rm GeV}}\right)
\end{equation}
in terms of the yield, where $n_{3/2}$ is the number density of
gravitino and $s$ is the entropy density. In the right hand side, $T_R$
is the reheating temperature, $m_{3/2}$ is the gravitino mass, $m_{Gi}$
is the gaugino masses for $i$-th generation, and $g_i$ is the gauge
coupling\footnote{
  In fact, the gauge coupling and the gaugino masses mildly depend on
  the reheating temperature $T_R$. However
  we neglect the weak dependences  in the following discussions.

}. Notice that the relic abundance is proportional to the reheating
temperature $T_R$. In the modulated reheating scenario, the reheating
temperature fluctuates in space, which implies that the number density
of gravitinos per entropy also fluctuates. Then, if the gravitinos
constitute DM of the Universe, isocurvature fluctuations are
generated. In order to discuss it more quantitatively, we relate the
fluctuations of the number density of gravitinos to the curvature
fluctuations as
\beq
S_{3/2} \equiv \frac{\delta(n_{3/2}/s)}{n_{3/2}/s}~.
  \label{eq:graiso1}
\eeq   
In the modulated reheating scenario, $\delta T_R / T_R \ne 0$, which
implies that a significant isocurvature fluctuation is generated as
\begin{equation}
S_{3/2} = \frac{\delta T_R}{T_R}. 
  \label{eq:graiso2}
\end{equation}
Since $T_R$ is proportional to $\Gamma^{1/2}$, we find $\delta T_R/T_R =
\delta\Gamma/(2 \Gamma) = \delta x / (2x)$.  Then using the expression
for the curvature perturbation~(\ref{modcurv}), $S_{3/2}$ can be related
to $\zeta$ as
\begin{equation}
  S_{3/2} =  \frac{\delta T_R}{T_R}
               = \frac{1}{2x Q'(x)} \zeta \simeq -3 \zeta,
               \label{eq:iso}
\end{equation}
when $x \ll 1$.  Thus we have $S_{3/2}/ \zeta \simeq -3$, which is
totally (anti)-correlated to the curvature perturbation and already
contradicts with the current observation~\cite{Komatsu:2008hk}.
%
%
%
This shows that a scenario with thermally produced gravitino DM cannot
be viable in the modulated reheating scenario, in which large
non-Gaussianity can be generated, because of too large isocurvature
fluctuations.

In fact, a similar argument also holds when gravitinos are produced
non-thermally from the decay of some heavy scalar field such as the
inflaton or moduli in the case they acquire a nonvanishing 
vacuum expectation value \cite{Nakamura:2006uc}, which 
dominates over the other channels \cite{also}.
In this case, the yield can
be written as
\begin{equation}
\label{eq:Yp_NT}
  \frac{n_{3/2}}{s} = \frac{3}{2} B_{3/2} \frac{T_R}{m_{3/2}},
\end{equation}
where $B_{3/2}$ is the branching ratio of the decay into gravitinos.
When gravitinos are produced from the jets, $B_{3/2}$ should be
understood as those including its multiplicity.  Since the reheating
temperature is related to the decay rate of the inflaton as $T_R \propto \Gamma^{1/2}$,
then $B_{3/2} \propto T_R^{-2}$, we have the
$T_R$-dependence of the $n_{3/2} / s$, in the case of non-thermal
production, as
\begin{equation}
 \frac{n_{3/2}}{s} \propto \frac{1}{T_R}.
\end{equation}
Thus isocurvature fluctuations are also generated when gravitinos are
produced from some heavy scalar decay, which can be written as
\begin{equation}
\label{eq:iso2}
S_{3/2} = - \frac{\delta T_R}{T_R},
\end{equation}
which is different in sign but the same in size as in the case that
gravitinos are produced thermally. Note that this isocurvature
fluctuation has totally {\it positive} correlation to the curvature
perturbation. Hence if DM consists of gravitinos, large isocurvature
fluctuations are generically produced in the modulated reheating
scenario, which is inconsistent with cosmological observations such as
CMB.

There is another well-known mechanism to generate large local-type
non-Gaussianity, called the curvaton mechanism. In this scenario, 
gravitinos are produced from the decays of both the
inflaton and the curvaton. As discussed  in
Ref.~\cite{Lyth:2002my,Lemoine:2009is}, 
in the former case, gravitino DM is disfavored
because of too large isocurvature fluctuations. On the other hand, in
the latter case, gravitino DM is permitted only when the curvaton almost
dominates the energy density of the universe to avoid large isocurvature
fluctuations, in which the non-Gaussianity is small. Then,
it is concluded that gravitino DM is disfavored also in the curvaton
scenario if the large non-Gaussianity is observed
\footnote{
For the case of baryonic isocurvature fluctuations, see Ref.~\cite{Moroi:2008nn}.
}. Therefore,
gravitino dark matter is disfavored once large local-type
non-Gaussianity of primordial fluctuations is observed.

Some comments are in order. First gravitinos can also be
produced by the decay of the next-to-the-lightest supersymmetric
particle (NLSP). However, the detailed calculations show that the
constraints from BBN are severer than those from the overclosure of the
universe, irrespective of the particle kind of NLSP as long as it is the
MSSM particle \cite{Kawasaki:2008qe}. Therefore, gravitino DM can be
realized only for the case that most of gravitinos should be produced
much earlier, either by the scattering processes during reheating or by
the non-thermal decay of heavy scalars, both of which we paid attention
to in this Letter.

Second, a similar argument also holds for the case where LSPs, which are
DM, are produced from the decay of gravitinos. In this case, LSPs carry
large isocurvature fluctuations originating from gravitinos, which again
contradicts with the present constraint. Thus, DMs originating from
gravitinos are also disfavored once large local-type non-Gaussianity is
detected\footnote{We thank M. Kawasaki and F. Takahashi for pointing
out this possibility.}.

Finally we comment on the case of axino DM. Axinos are produced by
thermal scattering during reheating and their abundance also depends on
the reheating temperature
\cite{Rajagopal:1990yx,Covi:2001nw,Covi:2002vw,Brandenburg:2004du}. Therefore,
as with the case of gravitinos, if such axinos become DM, they also
acquire large isocurvature fluctuations in the modulated reheating
scenario, which again contradicts with the present observational
constraints. Unlike the gravitino dark matter, however,
non-thermally produced axinos by the decay of NLSP
\cite{Covi:1999ty,Covi:2001nw,Brandenburg:2004du} or Q-balls
\cite{Roszkowski:2006kw} can be dominant with avoiding the BBN
constraints. For such cases, isocurvature constraints can be waived.

\bigskip
\bigskip

{\sl Acknowledgments:} 

We thank Takahiro Tanaka for the collaboration at the early stage. We
are also grateful to Takashi Hamazaki, Masahiro Kawasaki, Kazunori Kohri,
and Fuminobu Takahashi for useful discussions. This work is supported by
JSPS Grant-in-Aid for Scientific research, No.\,19740145 (T.T.),
No.\,19340054 (J.Y.), and No.\,21740187 (M.Y.).  S.Y. is supported in
part by Grant-in-Aid for Scientific Research on Priority Areas No. 467
``Probing the Dark Energy through an Extremely Wide and Deep Survey with
Subaru Telescope''.  He also acknowledges the support from the
Grand-in-Aid for the Global COE Program ``Quest for Fundamental
Principles in the Universe: from Particles to the Solar System and the
Cosmos '' from the Ministry of Education, Culture, Sports, Science and
Technology (MEXT) of Japan. We would like to thank the organizers of the
IPMU workshop on ``Focus week on non-Gaussianities in the sky'' and the
GCOE/YITP workshop YITP-W-09-01 on ``Non-linear cosmological
perturbations'' for their hospitality, during which a part of this work
is done.



\end{document}